\documentclass{jaa}
\usepackage{natbib}
\setcitestyle{aysep={}}

\usepackage{graphicx}
\usepackage{hyperref}
\hypersetup{colorlinks=true, allcolors=blue!60!black}
\usepackage{amsmath}
\usepackage{nccmath}
\usepackage{lastpage}
\usepackage{xcolor}
\newcommand{\rev}[1]{\textcolor{black}{#1}}

\newcommand{\Ipivot}{351.1}
\newcommand{\Cfit}{17.389}
\newcommand{\Cerr}{0.0043}
\newcommand{\Dfit}{1.392}
\newcommand{\Derr}{0.066}
\newcommand{\Crho}{+0.47}
\newcommand{\Sint}{0.104}
\newcommand{\Ncal}{605}
\newcommand{\Bfwd}{0.366}
\newcommand{\Bferr}{0.015}
\newcommand{\BDprod}{0.51}
\newcommand{\HoBias}{+0.052}
\newcommand{\HoRMS}{0.083}
\newcommand{\HoPct}{14}
\newcommand{\HoN}{272}
\newcommand{\Sone}{0.928}
\newcommand{\Stwo}{0.791}
\newcommand{\Sthree}{0.685}
\newcommand{\NaISevenThree}{181}
\newcommand{\NaINineTwo}{152}
\newcommand{\NaZeroNine}{137}
\newcommand{\NaThreeSeven}{135}
\newcommand{\MuLo}{0.66}
\newcommand{\MuHi}{0.95}
\newcommand{\StaLo}{168}
\newcommand{\StaHi}{172}
\newcommand{\MuResid}{-0.011}

\newcommand{\TotScatter}{0.095}
\newcommand{\MeanScatter}{0.104}
\newcommand{\AnchorSpread}{33--64}
\newcommand{\HoBiasMean}{+0.058}
\newcommand{\FsysEffect}{1.5}
\newcommand{\EnvLo}{13}
\newcommand{\EnvHi}{41}
\newcommand{\CtoL}{0.46}
\newcommand{\CtoLmax}{16}
\newcommand{\FarMuLo}{0.42}
\newcommand{\FarMuHi}{0.99}
\newcommand{\Cadence}{36~min}
\newcommand{\NblockC}{0.018}
\newcommand{\NblockD}{0.212}
\newcommand{\Neff}{58}
\newcommand{\MuInduced}{0.003}

\begin{document}\sloppy

\title{On Modelling the Far-Side Unsigned Magnetic Flux Evolution of Long-Lived NOAA Active Regions}

\author{AHAAN NALAVADI\textsuperscript{1,2,*} and MANJU SUDHAKAR\textsuperscript{2}}
\affilOne{$^1$Australian National University, Canberra, ACT 2611, Australia.\\}

\affilTwo{$^2$Space Astronomy Group, U. R. Rao Satellite Centre, ISRO, Bengaluru 560 037, India.}

\twocolumn[{

\maketitle

\corres{ahaan.nalavadi@gmail.com}

\msinfo{15 April 2026}{03 November 2026}

\begin{abstract}
The evolution of magnetic flux in solar active regions is central to both space-weather forecasting and solar-cycle studies. Direct measurements of the far-side magnetic field, however, remain unavailable with current solar instrumentation. In this paper, we develop a method to estimate the far-side evolution of active-region magnetic flux using 304~\AA\ (He~II) emission as a quantitative proxy. Using SDO/AIA and SDO/HMI observations of all four near-side passages of a long-lived active region (NOAA AR 12192), we fit a single relation that gives mean per-pixel unsigned magnetic flux from mean per-pixel 304~\AA\ intensity. The fit uses only disk-centre, flare-free intervals, and flux is the dependent variable, so the relation can be applied directly to intensity measurements. The magnetogram flux is deprojected to the radial direction before fitting, and we report the line-of-sight version for comparison. We first test the relation by fitting it on the first two passages and predicting the two later passages, which were not used in that fit; the relation adopted for the reconstruction is then fitted to all four passages. We then apply the same relation to STEREO-A/EUVI 304~\AA\ observations, scaling the EUVI intensities to AIA-equivalent boundary intensities inferred from the adjacent near-side flux measurements, and so reconstruct the unsigned flux history while the region was out of view from Earth. The test on the later passages gives a median log-space residual of \HoBias~dex, a root-mean-square error of \HoRMS~dex, and a median absolute fractional error of \HoPct\% over \HoN\ samples that were not used in the fit. Because a single cross-instrument scale factor cannot match both ends of a far-side interval, the absolute level of each reconstructed segment carries an uncertainty of several tens of per cent; it is the shape of the far-side curve, rather than its level, that the EUVI observations constrain.
\end{abstract}


}]

\year{2026}
\setcounter{page}{1}

\section{\rev{Introduction}}
\begin{figure*}[tp]
    \centering
    \includegraphics[width=\textwidth]{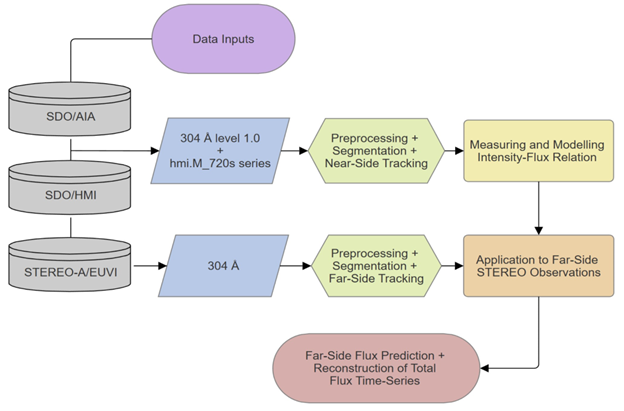}
    \caption{Schematic overview of the analysis pipeline used to reconstruct the far-side unsigned magnetic-flux evolution of the long-lived active region. The procedure begins with near-side calibration using SDO/AIA 304~\AA\ images and co-spatial SDO/HMI magnetograms, followed by active-region segmentation, fixed-area reprojection into a common CEA grid, and measurement of mean per-pixel intensity and unsigned magnetic flux. The calibrated relation is then applied to STEREO-A/EUVI 304~\AA\ observations, with the EUVI intensities scaled to AIA-equivalent boundary intensities inferred from the adjacent near-side flux measurements, giving the far-side flux history and the final stitched time series.}
    \label{fig:pipeline}
\end{figure*}

Solar activity is largely governed by the Sun's magnetic field. The dynamic behavior of this field drives key energetic phenomena such as solar flares and coronal mass ejections (CMEs), primarily in solar active regions (ARs). The frequency and intensity of these events fluctuate over an 11-year activity cycle governed by the solar dynamo, which itself is driven by differential rotation, meridional flows, and convective motions in the solar interior that generate, organize, and transport magnetic flux across the solar surface. The magnetic flux content and evolution in active regions thus govern both their eruptive potential and their contribution to the global solar magnetic field. Recent missions such as STEREO (2006) and SDO (2010) transformed solar monitoring by enabling sustained EUV observations of both the Earth-facing disk and, during favorable spacecraft geometry, much of the far side; this was particularly valuable while STEREO-B was still operational. SDO/HMI measures the photospheric magnetic field on the Earth-facing side, but no equivalent system provides continuous far-side magnetographic coverage. Although Solar Orbiter (2020) supplies off-Sun--Earth-line magnetic and EUV data through PHI and EUI, its orbit and operational constraints do not yield the continuous, global coverage required for routine space-weather forecasting. As a result, tracking active-region evolution before it rotates into terrestrial view remains challenging. During Solar Cycle 24, for example, it was estimated that on the order of $10^{22}$~Mx of flux ($25$--$30$\% of total emerging flux) originated on the far side of the Sun \citep{Upton2024}. As such, one solution for capturing this `missed' flux is the use of EUV emission in the 304~\AA\ (He II) line as a quantitative proxy. While ground-based chromospheric proxies, such as Ca~II~K, have long-established correlations with magnetic flux in ARs and plages \citep[and references therein]{Chatzistergos2019}, they are inherently restricted to the Earth-facing solar disk. The 304~\AA\ He~II line, measured by both SDO/AIA and STEREO/EUVI, provides a space-based equivalent for monitoring chromospheric and transition-region activity, offering the most reliable representation of these features for reconstructing magnetic flux evolution on the solar far side. \citet{UgarteUrra2015} showed that 304~\AA\ intensity can serve as a reliable proxy for magnetic flux, reporting a power-law relation across multiple ARs between the evolutions of magnetic flux from SDO/HMI and 304~\AA\ emissions from SDO/AIA. Already, 304~\AA\ observations have been used to assess the accuracy of helioseismic far-side imaging techniques. \citet{Liewer2012,Liewer2014,Liewer2017} compared STEREO/EUVI observations with helioseismic holography images and found that 95\% of helioseismically detected far-side ARs corresponded to 304~\AA\ brightenings. Similarly, \citet{Zhao2019} reported a 97.3\% match for time-distance helioseismology-detected far-side ARs above a certain size threshold. This rich literature underpins the rationale for using 304~\AA\ as a quantitative proxy for unsigned magnetic flux. \rev{Work using EUV 304~\AA\ emission as a proxy for solar magnetic activity falls into three groups: empirical proxy studies, which relate 304~\AA\ intensity directly to magnetic flux; machine-learning studies, which generate magnetograms or flux maps from EUV images; and operational studies, which feed far-side EUV observations into flux-transport or forecasting models.} \rev{The present work belongs to the first group, but aims at a quantitative flux estimate with stated uncertainties rather than at detecting active regions.} For example, \citet{Jeong2022} used deep learning to generate solar far-side magnetograms from STEREO/EUVI observations together with front-side SDO constraints, while \citet{Chen2022} used machine learning to infer magnetic-flux maps from 304~\AA\ and helioseismic observations. At a more operational level, \citet{Upton2024} used STEREO 304~\AA\ data to infer the emergence and unsigned flux of far-side active regions for ingestion into the Advective Flux Transport model. \rev{Unlike such approaches, the present work focuses on the quantitative evolution of a single long-lived active region, reconstructing its unsigned magnetic flux across multiple rotations using a calibrated 304~\AA\ proxy.}

Against this background, we explore whether the unsigned magnetic flux history of a single long-lived active region can be followed across repeated near- and far-side passages using EUV 304~\AA\ emission as a practical proxy. We combine SDO/AIA 304~\AA\ observations, SDO/HMI magnetograms, and STEREO/EUVI 304~\AA\ data to calibrate this proxy on the near side and then apply it to the corresponding far-side intervals of the same region. \rev{The aim is to obtain a consistent and interpretable estimate of the changes in unsigned flux over multiple solar rotations.} \rev{In brief, we: (1) identify and segment the LLAR in full-disk AIA and EUVI 304~\AA\ images; (2) track it using a fixed-area box of the same dimensions as an HMI SHARP patch, centred at each time step on the centroid of the segmented component; (3) reproject AIA, HMI and far-side EUVI onto the same CEA grid; (4) measure the total intensity, the number of segmented pixels and the total unsigned flux, and from these the mean per-pixel intensity and mean per-pixel flux; (5) fit a single relation giving flux from intensity, using only disk-centre, flare-free near-side data; (6) scale the EUVI intensities to AIA-equivalent boundary intensities, inferred from the adjacent near-side flux measurements through the same relation; and (7) apply the same relation to the scaled EUVI intensities to obtain the far-side flux history, and join it to the near-side measurements.} The overall workflow is summarized in Figure~\ref{fig:pipeline}. The observational data sets are introduced in the next section, followed by the analysis procedure.

\begin{figure*}[tp]
    \centering
    \includegraphics[width=\textwidth]{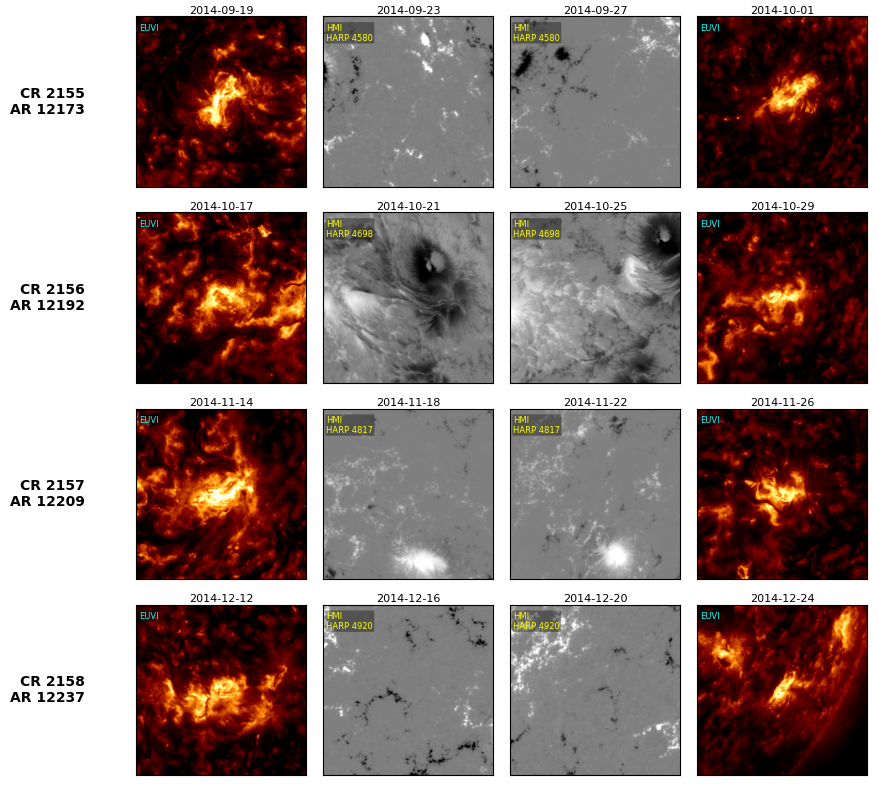}
    \caption{\rev{Representative multi-rotation context for the long-lived active region later identified most prominently as NOAA AR 12192.} For each Carrington rotation, the middle two panels show the region on the Earth-facing side in SDO/HMI line-of-sight magnetograms, while the outer panels show representative far-side appearances in STEREO-A/EUVI 304~\AA\ images. The figure is intended to illustrate the continuity of the same long-lived region across successive rotations using the observational datasets analysed in this work. \rev{The active-region identification information is summarized in Table~\ref{tab:lifetime}.}}
    \label{fig:lifetime}
\end{figure*}

\section{Observations}

We track AR 12192 across four Carrington rotations from 22 September 2014 to 21 December 2014 using data from the Atmospheric Imaging Assembly \citep[AIA;][]{Lemen2012} and the Helioseismic Magnetic Imager \citep[HMI;][]{Schou2012,Scherrer2012} onboard the Solar Dynamics Observatory \citep[SDO;][]{Pesnell2012}. We then apply the observed flux--intensity relation to far-side data from the Extreme Ultraviolet Imager \citep[EUVI;][]{Wuelser2004} on the Solar Terrestrial Relations Observatory Ahead spacecraft \citep[STEREO-A;][]{Kaiser2008}. 

\subsection{AR 12192 in multiple Carrington rotations}

\begin{table*}[tp]
\centering
\tabularfont
\caption{\rev{Near-side intervals used for AR 12192, with identification information, the range of $\mu$ spanned by the retained samples, and the STEREO-A--Earth separation angle at central meridian passage.}}
\label{tab:lifetime}
\begin{tabular}{lccccc}
\topline
\textbf{Dates} & \textbf{CR} & \textbf{NOAA AR} & \textbf{HARPNUM} & \boldmath$\mu$ \textbf{range} & \textbf{STEREO-A sep.}\\
\midline
Sept 22--29 & CR 2155 & 12173 & 4580 & 0.66--0.94 & $168^\circ$\\
Oct 18--27  & CR 2156 & 12192 & 4698 & 0.70--0.94 & $170^\circ$\\
Nov 15--22  & CR 2157 & 12209 & 4817 & 0.66--0.95 & $171^\circ$\\
Dec 13--21  & CR 2158 & 12237 & 4920 & 0.73--0.95 & $172^\circ$\\
\hline
\end{tabular}
\tablenotes{\rev{\textbf{Note.} Dates give the interval over which the region lay within $0.9\,R_\odot$ of disk centre; data outside this radius were discarded, so the dates do not cover the whole passage across the visible disk. $\mu$ is the cosine of the heliocentric angle, averaged over the segmented mask.} \rev{The separation angle is the Stonyhurst longitude of STEREO-A, computed from the spacecraft ephemeris.}}
\end{table*}

The National Oceanic and Atmospheric Administration (NOAA) assigns unique numerical identifiers to ARs when they first appear on the visible solar disk. Every time an active region rotates behind the Sun and re-emerges on the eastern limb, NOAA's system assigns it a new AR number. As a result, any long-lived active region like AR 12192 can undergo multiple `name' changes during its lifetime. Figure~\ref{fig:lifetime} therefore provides representative multi-rotation context for the same LLAR using the observational datasets analysed in this work: SDO/HMI line-of-sight magnetograms on the Earth-facing side and STEREO-A/EUVI 304~\AA\ images on the far side. Each row corresponds to a new Carrington rotation (CR) and therefore a new NOAA AR number assigned to the active region. The numbers assigned to AR 12192 over its lifetime are listed in Table~\ref{tab:lifetime} alongside other relevant identification information.

\subsection{Solar Dynamics Observatory (SDO)}

SDO operates in a geosynchronous inclined orbit with a perspective along the Sun--Earth line. SDO/AIA provides full-disk, EUV and UV images in multiple passbands, enabling high-cadence (12~s) studies of the solar chromosphere, transition region, and corona. \rev{We use He~II 304~\AA\ images at a spatial resolution of 0.6 arcsec per pixel, downloaded with SunPy's Fido from the \texttt{aia.lev1\_euv\_12s} series and processed to Level-1.5 with standard SunPy and \texttt{aiapy} routines.} \rev{Processing comprises the pointing correction, registration to a common 0.6 arcsec per pixel plate scale with solar north up, correction for the time-dependent degradation of the 304~\AA\ channel, and division by the exposure time.} \rev{All intensities quoted in this paper are therefore in data numbers per second (DN~s$^{-1}$).} \rev{Exposure times in the analysed intervals are constant to better than 0.05\%, so this normalisation acts as a fixed rescaling and does not affect the scatter of the calibration.}

\rev{SDO/HMI provides full-disk line-of-sight and vector magnetograms at 0.5 arcsec per pixel with a cadence of 45~s.} \rev{We use the \texttt{hmi.M\_720s} line-of-sight magnetograms provided by the Joint Science Operations Center\footnote{\url{http://jsoc.stanford.edu}}, treating the line-of-sight field as an unsigned quantity.} \rev{Because the line-of-sight component under-estimates the field of a predominantly radial active-region flux system by a factor $\mu$, we also form a deprojected estimate $B_r \simeq B_{\mathrm{LOS}}/\mu$ and report both; this is discussed in Section~\ref{sec:geometry}.} \rev{AIA and HMI frames are paired on nearest observation time, with pairs separated by more than 6~min discarded.}

\subsection{Solar Terrestrial Relations Observatory (STEREO)}

The STEREO mission consists of two spacecraft placed in heliocentric orbits ahead of and behind Earth, enabling stereoscopic and far-side observations of the Sun. STEREO-A/EUVI observes the solar disk in four EUV passbands with a spatial resolution of 1.6 arcsec per pixel and a nominal cadence of 10~min at 304~\AA. Near the end of 2014, STEREO-A viewed AR 12192 as it rotated across the far side of the Sun and back into view. \rev{We use STEREO-A/EUVI 304~\AA\ images to infer the far-side unsigned-flux evolution in AR 12192.}

\subsection{\rev{Viewing geometry and projection effects}}
\label{sec:geometry}

\rev{The fixed-area CEA reprojection described in Section~3 removes the geometric foreshortening of \emph{area} from both $\bar{I}$ and $\bar{\Phi}$ by construction, because each grid pixel subtends a constant physical area on the solar surface irrespective of where the region lies on the disk.} \rev{Two effects survive this reprojection and are treated explicitly.}

\rev{First, the line-of-sight magnetic field is projected.} \rev{Over the $\pm 3$-day calibration window the mask-averaged $\mu$ ranges from \MuLo\ to \MuHi, so $|B_{\mathrm{LOS}}|$ under-estimates $|B_r|$ by up to a few tens of per cent, and the size of the shortfall is correlated with $\mu$.} \rev{We therefore report the calibration for both the line-of-sight and the deprojected flux, and adopt the deprojected version as the primary result.} \rev{The deprojection assumes that the active-region field is predominantly radial, which is a good approximation for the strong-field interior of an active region but less so near its periphery.}

\rev{Second, the 304~\AA\ radiance itself varies from disk centre to limb.} \rev{Comparing the mask-averaged intensity between disk-centre and near-limb intervals of the same region gives $u = \CtoL$ in $I(\mu) = I_0[1 - u(1-\mu)]$, which would imply a variation of up to \CtoLmax\% at the edge of the calibration window, about twice the systematic term of Equation~(\ref{eq:sigma_i}).} \rev{That figure should not be read as a limb-darkening law: the comparison cannot separate the region's decay between the two epochs from the geometric variation, so it is a loose upper bound and no more.} \rev{A direct test is more informative.} \rev{A centre-to-limb dependence of this size would appear as a trend in the calibration residuals with $\mu$; the measured trend is \MuResid~dex per unit $\mu$ (Section~\ref{sec:sensitivity}), implying a contribution of \MuInduced~dex across the retained range, more than an order of magnitude below the intrinsic scatter of \Sint~dex.} \rev{We therefore apply no correction, relying on the residual test rather than on $u$.} \rev{A related effect is that near the limb the region is sampled by fewer detector pixels, so the reprojection smooths the mask boundary slightly; this is confined to the edges of the calibration window.}

\rev{For the far side, STEREO-A led Earth by \StaLo$^\circ$ in late September 2014 and \StaHi$^\circ$ by late December 2014.} \rev{At these separations the centre of the terrestrial far side lay approximately $8^\circ$--$12^\circ$ from the centre of the STEREO-A disk, so STEREO-A provided substantially more favourable viewing geometry for these intervals than any Earth-based vantage point.} \rev{Over the far-side samples retained, $\mu$ in the STEREO-A frame ranges from \FarMuLo\ to \FarMuHi; samples below $\mu = 0.4$ are excluded.}

\section{Active region tracking and segmentation}

\begin{figure*}[tp]
    \centering
    \includegraphics[width=0.9\textwidth]{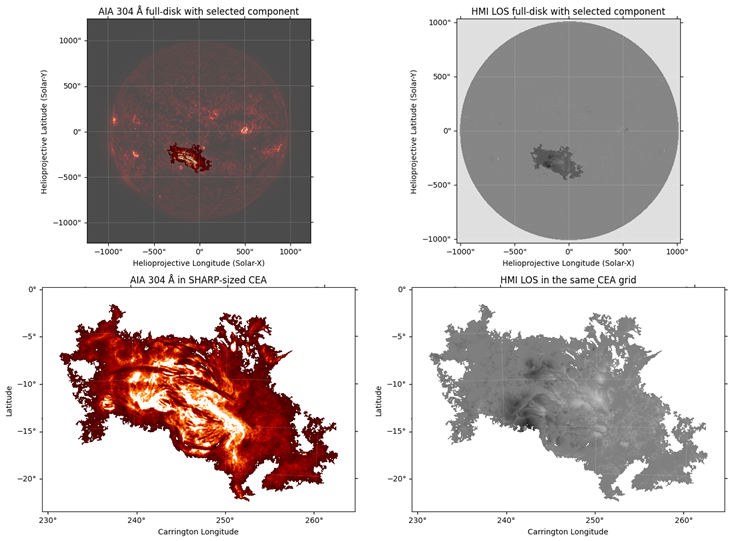}
    \caption{\rev{Representative example of the intensity-based segmentation and reprojection procedure used in this study.} The top-left panel shows the full-disk SDO/AIA 304~\AA\ image from which the long-lived active region is identified, while the top-right panel shows the corresponding region in the full-disk SDO/HMI line-of-sight magnetogram. The bottom panels show the same selected region after reprojection into the common fixed-area cylindrical equal-area (CEA) grid used for subsequent measurements. \rev{This common grid ensures that the intensity and flux are evaluated over the same physical area, and removes first-order changes associated with projected area; residual line-of-sight and centre-to-limb effects are treated separately in Section~\ref{sec:geometry}.}}
    \label{fig:segmentation_nearside}
\end{figure*}

\rev{For each AIA image, we compute the median intensity over the solar disk and identify pixels exceeding:
\begin{equation}
I(x,y) > k \times \mathrm{median}(I)
\label{eq:threshold}
\end{equation}
with $k=3$ in Equation~(\ref{eq:threshold}).} This simple thresholding approach is widely used in EUV-based detection of active regions and plages and has been shown to robustly isolate large-scale magnetic structures in 304~\AA\ emission \citep{Liewer2012,UgarteUrra2015}. To suppress noise and small-scale transient brightenings, the resulting binary mask is subjected to morphological opening and closing operations. Connected components are then identified, and only regions exceeding a minimum area threshold of 500 pixels are retained. When multiple regions are present, the largest connected component is selected, consistent with the expectation that the LLAR dominates the 304~\AA\ emission during the analysed intervals. For each detected active region, a fixed-size cylindrical equal-area (CEA) map is constructed, centred on the Carrington longitude and latitude of the region's intensity-weighted centroid. The map dimensions are chosen to match those of an HMI SHARP CEA patch, ensuring a constant physical area is sampled throughout the region's evolution. Both the AIA 304~\AA\ image and the corresponding HMI magnetogram are reprojected onto this synthetic CEA grid and thus for each AIA--HMI data pair, the exact same pixels used for AIA intensity calculation are used for the flux calculation. \rev{This removes, to first order, the changes that would otherwise arise from the projected area of the region and from the physical area sampled.} \rev{Residual projection effects on the line-of-sight field and on the 304~\AA\ radiance survive this step and are treated separately in Section~\ref{sec:geometry}.} Similar fixed-area reprojection strategies are commonly used in studies of active region magnetic evolution \citep{Bobra2014,Leka2018}. \rev{Figure~\ref{fig:segmentation_nearside} shows the segmentation applied to one AIA image and the corresponding HMI magnetogram.}

\rev{Samples are drawn at a uniform cadence of \Cadence, rather than at the $\sim$12-min cadence at which the HMI 720-s series and the AIA images used here are sampled.} \rev{This is a practical sampling choice and not a statistical one: successive frames remain correlated at either cadence, and that correlation is addressed directly in Section~\ref{sec:sensitivity} rather than by subsampling.} 

\rev{Beyond the segmentation itself, several criteria are applied before a sample enters the calibration.} \rev{These are listed in Table~\ref{tab:criteria} for both the near-side and the far-side data.} \rev{Two of them differ between the two sets: the disk-centre window applies only to the near side, where central meridian passage is defined, and a $2\sigma$ clip on $\log_{10}\bar{I}$ is applied only to the far side, where isolated intensity spikes cannot be cross-checked against a magnetogram.} \rev{The remaining criteria are common to both.}

\begin{table}[tbp]
\centering
\tabularfont
\caption{\rev{Selection criteria applied to near-side and far-side samples.}}
\label{tab:criteria}
\begin{tabular}{lcc}
\topline
\textbf{Criterion} & \textbf{Near side} & \textbf{Far side}\\
\midline
Threshold multiplier $k$        & 3      & 3\\
Minimum mask area (pixels)      & 500    & 500\\
Largest connected component     & yes    & yes\\
Disk-centre window              & $\pm 3$~d & ---\\
Flare exclusion (GOES/HEK)      & yes    & ---\\
Max.\ centroid longitude step   & $8^\circ$ & $8^\circ$\\
Latitude clip                   & $4\sigma$ & $4\sigma$\\
Minimum contiguous run          & 3      & 3\\
$\log_{10}\bar{I}$ spike clip    & ---    & $2\sigma$, 2 iter.\\
STEREO-A viewing angle           & ---    & $\mu \ge 0.4$\\
AIA--HMI pairing tolerance      & 6~min  & ---\\
Sampling cadence                & \Cadence & as available\\
\hline
\end{tabular}
\tablenotes{\rev{\textbf{Note.} The longitude-step, latitude-clip and minimum-run criteria reject frames in which the tracked centroid jumps, which occurs when the segmentation momentarily latches onto a neighbouring brightening.}}
\end{table}

\begin{figure*}[tp]
    \centering
    \includegraphics[width=0.9\textwidth]{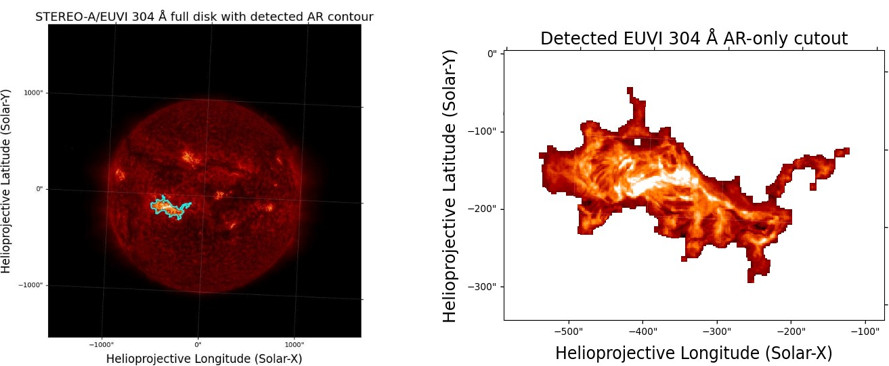}
    \caption{\rev{Representative example of the segmentation and reprojection procedure on a far-side STEREO-A/EUVI 304~\AA\ image from the days leading up to the 2014 October 14 eruptive event associated with this AR.} \rev{The right panel shows the selected region after reprojection into the common fixed-area cylindrical equal-area (CEA) grid used for measurements.}}
    \label{fig:segmentation_farside}
\end{figure*}

\begin{figure*}[tp]
    \centering
    \includegraphics[width=\textwidth]{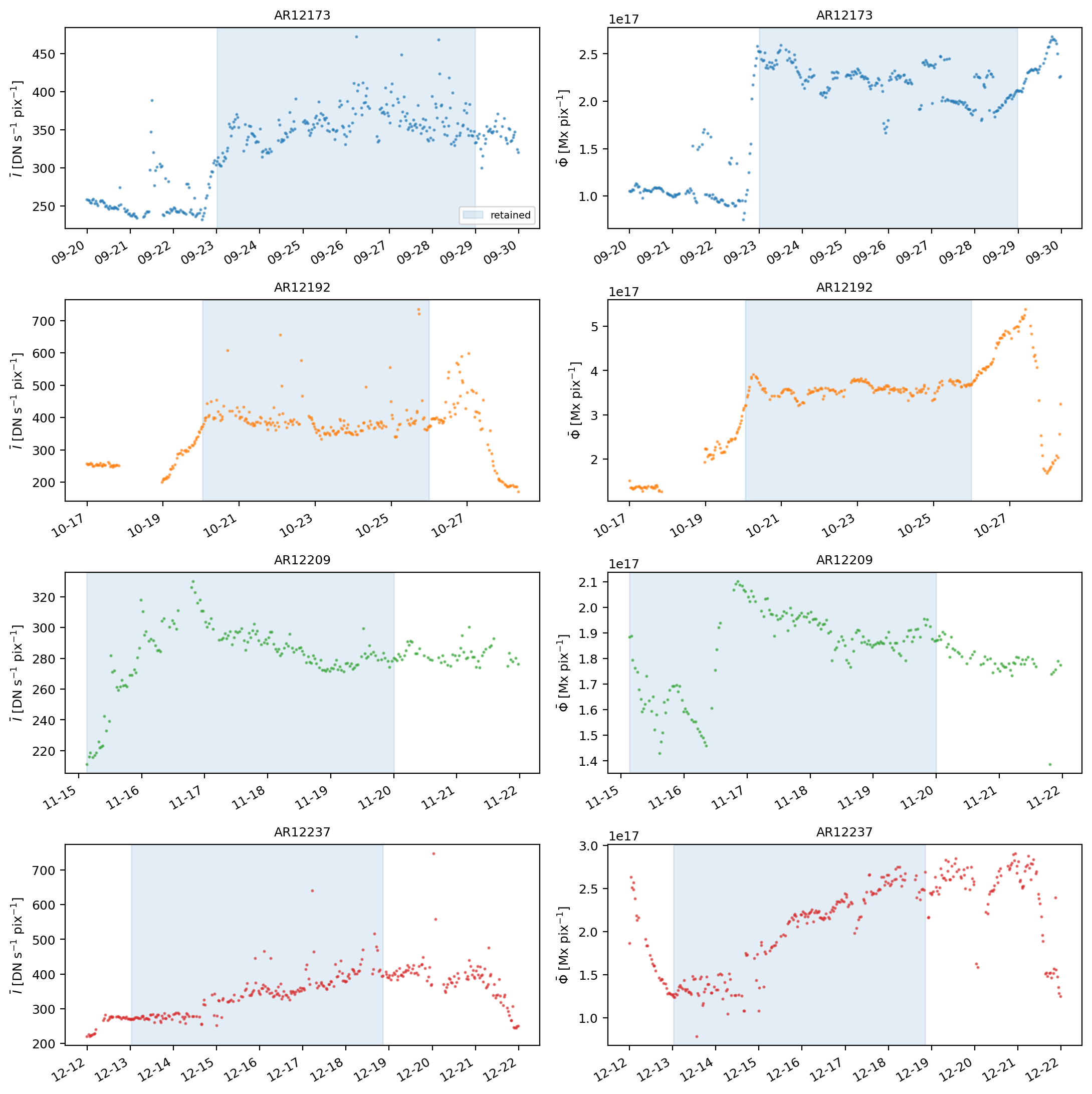}
    \caption{\rev{Time series of mean per-pixel 304~\AA\ intensity (top) and mean per-pixel unsigned flux (bottom) for each near-side passage of the long-lived active region.} \rev{All samples surviving segmentation are plotted; the shaded interval marks the window retained for the near-side calibration after the disk-centre, flare and tracking criteria of Table~\ref{tab:criteria} are applied.} \rev{The widths of the shaded intervals differ between passages because the retained samples reflect not only the nominal disk-centre window but also data availability and tracking quality.} \rev{For clarity, pointwise uncertainty bars are not shown; the uncertainty model used in the calibration and the reconstruction is described in Section~4.2.}}
    \label{fig:timeseries}
\end{figure*}

\section{Far-Side Unsigned Magnetic Flux Estimation}

\subsection{\rev{Mean per-pixel intensity and flux measurement}}

Within the reprojected active region mask, we compute the total 304~\AA\ intensity and total unsigned flux. \rev{Mean per-pixel quantities are then defined by Equation~(\ref{eq:means}),
\begin{equation}
\bar{I} = \frac{1}{N}\sum_{i=1}^{N} I_i,\qquad
\bar{\Phi} = \frac{1}{N}\sum_{i=1}^{N} |B_i|\,\Delta A,
\label{eq:means}
\end{equation}
where $N$ is the number of pixels in the segmented region, $I_i$ is the exposure-normalised AIA 304~\AA\ intensity of pixel $i$ in DN~s$^{-1}$, $B_i$ is the HMI magnetic field strength in Gauss (either the line-of-sight field or its deprojected form, as described in Section~\ref{sec:geometry}), and $\Delta A$ is the pixel area in the CEA projection.} 

Using mean per-pixel quantities rather than total values removes first-order dependence on region size and enables a direct comparison between intensity and flux density. This formulation is consistent with previous studies examining EUV--magnetic field relationships in active regions \citep{UgarteUrra2015,Madjarska2021}.

\subsection{Uncertainties}

Uncertainties in the total unsigned flux are estimated assuming uncorrelated pixel-scale noise in the HMI LOS magnetograms. The flux uncertainty is given by
\begin{equation}
\sigma_{\Phi} = \Delta A\,\sigma_B\sqrt{N},
\label{eq:sigma_phi}
\end{equation}
where $\sigma_B = 6$~G is the nominal per-pixel noise level for HMI 720~s magnetograms and $N$ is the number of pixels in the active region mask. This prescription follows standard error propagation used in earlier analyses of HMI unsigned-flux measurements \citep{Liu2012,Bobra2014}. \rev{When the deprojected flux of Section~\ref{sec:geometry} is used, the same propagation applied to $|B_{\mathrm{LOS}}|/\mu$ gives $\sigma_\Phi = \Delta A\,\sigma_B\left(\sum_i \mu_i^{-2}\right)^{1/2}$, which is the form used for the results reported here.} 

The uncertainty associated with extreme-ultraviolet (EUV) intensity measurements arises from multiple sources. While photon counting statistics and read noise affect individual pixels, these contributions become negligible when averaging over the large number of pixels ($N \gtrsim 500$) contained within an active region mask. 

Consequently, photon noise does not dominate the uncertainty in the mean per-pixel intensity $\bar{I}$ used in this study. Instead, the dominant sources of uncertainty are systematic in nature and include instrumental calibration effects, flatfield errors, uncertainties in the degradation correction, and variations in the temperature response of the 304~\AA\ channel. To account for these effects, we model the uncertainty in the mean EUV intensity as
\begin{equation}
\sigma_{\bar{I}}^2 = \left(\frac{\sigma_{\mathrm{pix}}}{\sqrt{N}}\right)^2 + \left(f_{\mathrm{sys}}\bar{I}\right)^2,
\label{eq:sigma_i}
\end{equation}
where $\sigma_{\mathrm{pix}}$ is an effective per-pixel noise scale and $f_{\mathrm{sys}}$ a fractional systematic uncertainty. \rev{We adopt $\sigma_{\mathrm{pix}} = 5$~DN~s$^{-1}$ per pixel and $f_{\mathrm{sys}} = 0.08$ for SDO/AIA, and $\sigma_{\mathrm{pix}} = 8$~DN~s$^{-1}$ per pixel and $f_{\mathrm{sys}} = 0.12$ for STEREO-A/EUVI.}

\rev{These values are deliberately conservative envelopes rather than formally propagated calibration errors.} \rev{The 304~\AA\ channel of AIA shows substantial time-dependent degradation, and the EVE-based correction for it leaves residuals at the several-per-cent level \citep{Boerner2012,Boerner2014}; the adopted 8\% is intended to cover that residual together with flat-field error and variation in the temperature response of the channel.} \rev{For EUVI we adopt a larger conservative envelope of 12\% to account additionally for the uncertainty of the EUVI-to-AIA cross-calibration.} \rev{Because these choices are not derived from first principles, we quantify in Section~\ref{sec:sensitivity} how much the results depend on them.} Because the mean flux uncertainty is derived from pixel-scale HMI noise that decreases as $\sim N^{-1/2}$ when averaged over the active-region mask, the resulting fractional uncertainty in $\bar{\Phi}$ is typically very small. \rev{By contrast, the uncertainty in $\bar{I}$ includes an explicit fractional systematic term, $f_{\mathrm{sys}}\bar{I}$, which does not average away with increasing mask size and therefore dominates the intensity uncertainty budget.}

\subsection{Near-side calibration}

The calibration is restricted to disk-centre intervals defined as $\pm 3$ days around each near-side central meridian passage. \rev{The intensity and flux time series for each AR are shown in Figure~\ref{fig:timeseries}, where the shaded interval marks the retained calibration window.} \rev{After all of the criteria in Table~\ref{tab:criteria} are applied, $n=\NaISevenThree$, \NaINineTwo, \NaZeroNine\ and \NaThreeSeven\ samples remain for AR 12173, 12192, 12209 and 12237 respectively.} 

\begin{figure}[tbp]
    \centering
    \includegraphics[width=\linewidth]{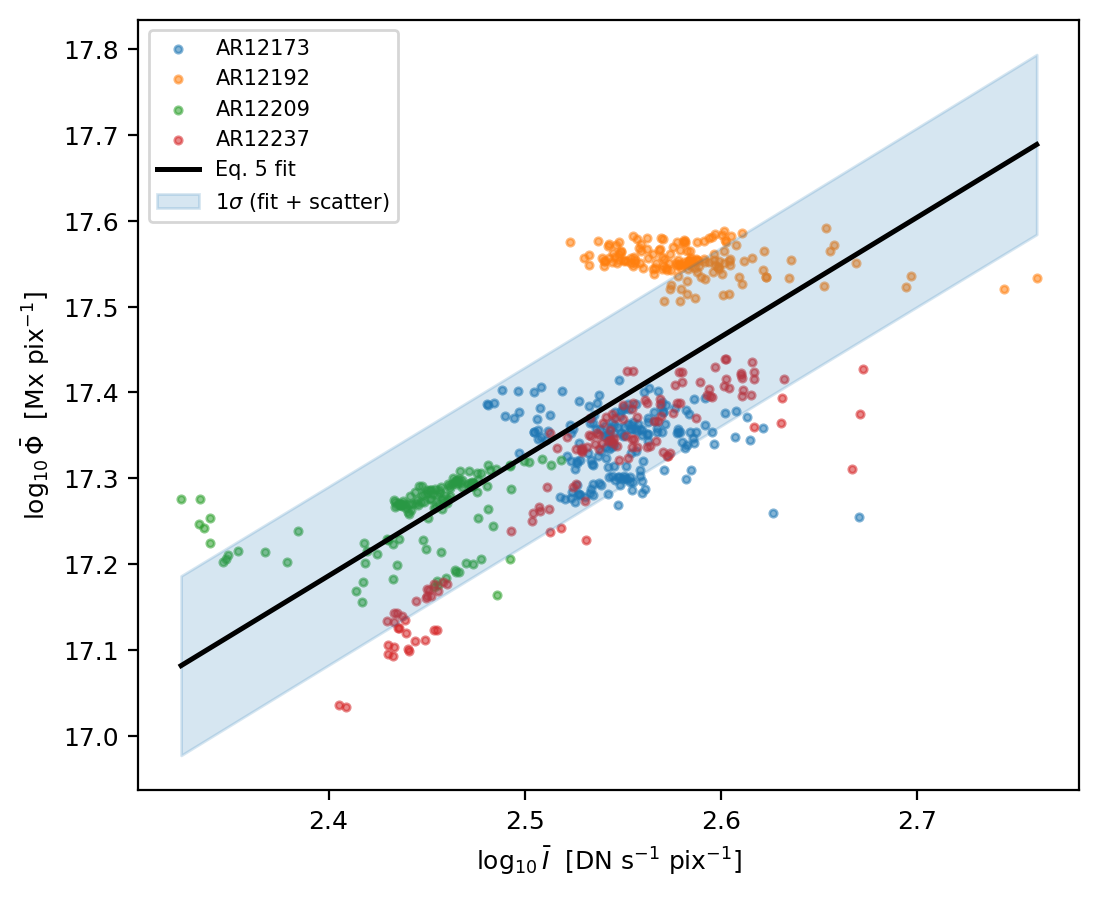}
    \caption{\rev{The calibration relation between mean per-pixel 304~\AA\ intensity and mean per-pixel unsigned flux, for the near-side sample.} \rev{The solid line is the best-fitting relation of Equation~(\ref{eq:inverse}), which is used everywhere in this work: to estimate far-side flux from EUVI intensity, and, inverted, to set the boundary anchors for the EUVI-to-AIA scaling.} \rev{The shaded band is the $1\sigma$ envelope of the fit in log space, from the fitted-parameter covariance together with the intrinsic scatter.}}
    \label{fig:inverse}
\end{figure}

\rev{We use a single relation throughout, fitted with flux as the dependent variable:
\begin{equation}
\log_{10}\bar{\Phi} = C + D\log_{10}\!\left(\frac{\bar{I}}{\bar{I}_0}\right),
\label{eq:inverse}
\end{equation}
where $\bar{I}_0 = \Ipivot$~DN~s$^{-1}$ per pixel is the median intensity of the calibration sample.} \rev{Referencing the intensity to $\bar{I}_0$ leaves the relation unchanged but substantially reduces the correlation between the fitted intercept and slope; without it, an intercept extrapolated to $\log_{10}\bar{I}=0$ lies far outside the data and is strongly anti-correlated with the slope.} \rev{The residual correlation of the two coefficients is given in Table~\ref{tab:coeffs}.} \rev{The fitted relation is shown in Figure~\ref{fig:inverse} and its coefficients are given in Table~\ref{tab:coeffs}.} 

\rev{Figure~\ref{fig:inverse} also shows that the four passages do not lie on a single tight locus, AR 12192 in particular sitting above the fitted line, so part of the relation is carried by the offsets between passages rather than by the correlation within any one of them.} \rev{This is why the formal parameter uncertainties are optimistic, as quantified by the block bootstrap of Section~\ref{sec:sensitivity}.}

\rev{The fit is a weighted least-squares fit in log space.} \rev{Weights come from the flux uncertainty of Equation~(\ref{eq:sigma_phi}), propagated into $\log_{10}\bar{\Phi}$, together with an intrinsic-scatter term added in quadrature and solved for so that the reduced $\chi^2$ of the fit equals unity.} \rev{The inverse of the weighted normal-equation matrix gives the formal parameter uncertainties, but these assume independent samples; the uncertainties quoted in Table~\ref{tab:coeffs} are instead those of the block bootstrap described in Section~\ref{sec:sensitivity}.}

\rev{Fitting with flux as the dependent variable matters because the two variables are not perfectly correlated.} \rev{For comparison, a fit made in the opposite direction has slope $B = \Bfwd \pm \Bferr$; if the two fits were exact inverses of one another the product $BD$ would equal unity, whereas in fact $BD = \BDprod$.} \rev{Inverting the forward fit to estimate flux would therefore give systematically different answers from Equation~(\ref{eq:inverse}), and the discrepancy is large, not marginal.} \rev{For this reason the forward fit is quoted here only for comparison with the power-law slopes reported by earlier proxy studies \citep{UgarteUrra2015}, and is not used anywhere in the reconstruction.}

\begin{table}[tbp]
\centering
\tabularfont
\caption{\rev{Coefficients of the calibration relation, Equation~(\ref{eq:inverse}), with $\bar{I}_0 = \Ipivot$~DN~s$^{-1}$ per pixel.}}
\label{tab:coeffs}
\begin{tabular}{lc}
\topline
\textbf{Quantity} & \textbf{Value}\\
\midline
Intercept $C$                         & $17.389 \pm \NblockC$\\
Slope $D$                             & $\Dfit \pm \NblockD$\\
Correlation of $C$ and $D$ (bootstrap) & \Crho\\
Intrinsic scatter (dex)               & \Sint\\
Sample size $n$                       & \Ncal\\
Reduced $\chi^2$                      & 1 (by construction)\\
Forward slope $B$, for comparison only & $\Bfwd \pm \Bferr$\\
Product $BD$                          & \BDprod\\
\hline
\end{tabular}
\tablenotes{\rev{\textbf{Note.} The uncertainties on $C$ and $D$ are from the one-day block bootstrap of Section~\ref{sec:sensitivity}.} \rev{The formal covariance-matrix values, which assume independent samples, are $\Cerr$ and $\Derr$ respectively; because consecutive observations are correlated these are optimistic and are not the values to carry.} \rev{The intrinsic scatter is solved for so that the reduced $\chi^2$ equals unity, so that value is not an independent goodness-of-fit statistic.} \rev{The forward slope $B$ is listed for comparison with earlier proxy studies only; $BD$ would equal unity if the forward and predictive fits were exact inverses.}}
\end{table}

\subsection{EUVI-to-AIA Scaling}

Far-side observations are obtained from the STEREO-A/EUVI instrument at 304~\AA. Active region segmentation is performed using the same intensity-based thresholding method as that used for AIA. A representative example of the far-side segmentation and reprojection procedure is shown in Figure~\ref{fig:segmentation_farside}. 

For clarity, we refer to the three far-side intervals as FS1, FS2, and FS3 (i.e.\ FS1 denotes the far-side interval between the near-side passages identified as AR 12173 and AR 12192, and so on). \rev{Because EUVI and AIA intensities are not directly comparable in absolute units, the EUVI intensities are multiplied by a factor $S$ chosen so that they match the AIA-equivalent intensities at the start and end of each far-side interval.} \rev{These are not measured AIA intensities, which do not exist for a region on the far side, but the AIA intensities implied by the adjacent near-side flux measurements through the calibration relation itself.} 

\rev{The AIA intensity expected at a boundary is obtained by solving Equation~(\ref{eq:inverse}) for $\bar{I}$,
\begin{equation}
\log_{10}\!\left(\frac{\hat{I}_{\mathrm{AIA}}}{\bar{I}_0}\right) = \frac{\log_{10}\bar{\Phi}_{\mathrm{HMI}} - C}{D},
\label{eq:anchor}
\end{equation}
where $\bar{\Phi}_{\mathrm{HMI}}$ is the measured near-side flux averaged over the 24~h adjacent to the boundary.} \rev{Because Equation~(\ref{eq:anchor}) is the exact algebraic inverse of Equation~(\ref{eq:inverse}), applying a boundary's own scale factor returns the measured near-side flux identically:
\begin{equation}
\begin{split}
\hat{\Phi} &= 10^{\,C + D\log_{10}(S\bar{I}_{\mathrm{EUVI}}/\bar{I}_0)}\\[2pt]
           &= 10^{\,C + D\left(\log_{10}\bar{\Phi}_{\mathrm{HMI}} - C\right)/D}
            = \bar{\Phi}_{\mathrm{HMI}},
\end{split}
\label{eq:continuity}
\end{equation}
Equation~(\ref{eq:continuity}) holds separately at each boundary, for that boundary's own scale factor: $S_{\mathrm{L}}$ at the ingress boundary and $S_{\mathrm{R}}$ at the egress boundary.} \rev{We adopt a single scale factor for each far-side interval, taken as the geometric mean $S = (S_{\mathrm{L}}S_{\mathrm{R}})^{1/2}$ so that neither boundary is privileged.} 

\rev{Unless $S_{\mathrm{L}} = S_{\mathrm{R}}$, this adopted $S$ reproduces neither boundary exactly and a residual offset remains.} \rev{Table~\ref{tab:scale} lists $S_{\mathrm{L}}$, $S_{\mathrm{R}}$, the adopted $S$, and the resulting mismatch at each boundary.} \rev{Substituting the fitted coefficients into Equation~(\ref{eq:inverse}) gives the relation used for the far-side reconstruction:}
\begin{equation}
\log_{10}\bar{\Phi} = \Cfit + \Dfit\,\log_{10}\!\left(\frac{S\bar{I}_{\mathrm{EUVI}}}{\bar{I}_0}\right).
\label{eq:final}
\end{equation}

\rev{It follows that the absolute level of the reconstructed far-side flux is set by the near-side measurements at the boundaries, to within the mismatch of Table~\ref{tab:scale}, and is not independent information.} \rev{What the EUVI observations supply is the \emph{shape} of the curve between those boundaries.}

\begin{table*}[tp]
\centering
\tabularfont
\caption{\rev{Boundary scale factors, the adopted geometric mean, and the resulting flux mismatch at each boundary of the three far-side intervals.}}
\label{tab:scale}
\begin{tabular}{lccccc}
\topline
\textbf{Passage} & \boldmath$S_{\mathrm{L}}$ & \boldmath$S_{\mathrm{R}}$ & \textbf{Adopted} \boldmath$S$ & \textbf{L mismatch} & \textbf{R mismatch}\\
\midline
FS1 (12173$\to$12192) & 0.777 & 1.108 & \Sone   & $+28.0$\% & $-21.8$\%\\
FS2 (12192$\to$12209) & 0.876 & 0.714 & \Stwo   & $-13.2$\% & $+15.2$\%\\
FS3 (12209$\to$12237) & 0.783 & 0.598 & \Sthree & $-17.1$\% & $+20.6$\%\\
\hline
\end{tabular}
\tablenotes{\rev{\textbf{Note.} $S_{\mathrm{L}}$ and $S_{\mathrm{R}}$ are the scale factors that reproduce the measured near-side flux exactly at the ingress and egress boundaries respectively; $S$ is their geometric mean, which is the value adopted.} \rev{The mismatch columns give the resulting fractional difference between the reconstructed and the measured flux at each boundary.}}
\end{table*}

\section{Results and Discussion}

\subsection{Validation Testing}

\begin{figure}[tbp]
    \centering
    \includegraphics[width=\linewidth]{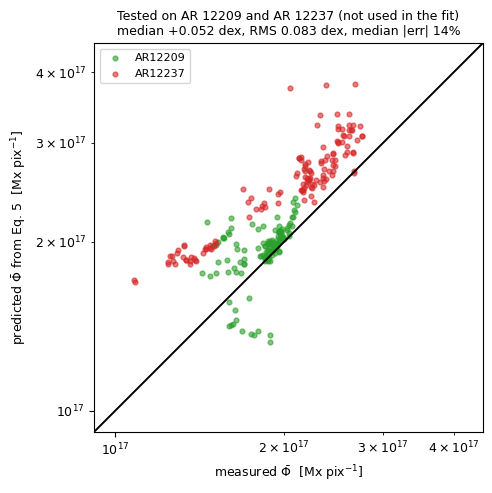}
    \caption{\rev{Test of Equation~(\ref{eq:inverse}) on passages not used in the fit.} \rev{The relation was fitted using the first two near-side passages, AR 12173 and AR 12192, and then applied to the two later passages, AR 12209 and AR 12237.} \rev{The horizontal axis is the flux measured from the magnetogram; the vertical axis is the flux predicted from the observed 304~\AA\ intensity.} \rev{The solid diagonal line indicates equality.}}
    \label{fig:validation}
\end{figure}

\rev{To test whether the relation holds beyond the data used to fit it, we fitted Equation~(\ref{eq:inverse}) using only the first two near-side passages, AR 12173 and AR 12192, and then used it to predict the flux for the two later passages, AR 12209 and AR 12237.} \rev{For each sample in the later passages the measured mean 304~\AA\ intensity was converted into a predicted flux and compared with the flux measured from the magnetogram for the same sample.} \rev{Figure~\ref{fig:validation} shows the comparison.} \rev{The fit used for this test is weighted in the same way as Equation~(\ref{eq:inverse}) itself, so that the relation being tested is the one actually being used.}

\rev{Over \HoN\ samples from the later passages the median log-space residual is \HoBias~dex, the root-mean-square residual is \HoRMS~dex, and the median absolute fractional error is \HoPct\%.} \rev{We quote the median rather than the mean residual because the residual distribution is not symmetric; the mean residual is \HoBiasMean~dex.} \rev{The relation therefore carries over to passages that were not used in fitting it, at a level adequate for far-side reconstruction, subject to the further systematic uncertainty of the EUVI-to-AIA scaling.}

\newpage \rev{The residual offset is worth noting explicitly.} \rev{A median residual of \HoBias~dex means the later passages are predicted systematically high, which could reflect either drift in the degradation correction for the AIA 304~\AA\ channel over the three months spanned by the four passages, or a real change in the ratio of 304~\AA\ emission to unsigned flux as the region decays.} \rev{The present data cannot separate the two, and we treat the offset as part of the quoted uncertainty rather than correcting for it.}

\subsection{\rev{Sensitivity Tests}}
\label{sec:sensitivity}

We perform a series of sensitivity tests before reconstructing the continuous flux time series. \rev{First, we test whether the relation depends on the use of mean per-pixel quantities by repeating the calibration using total EUV intensity and total unsigned flux.} \rev{A power-law relation is recovered in both cases, and the scatter is comparable: \TotScatter~dex using total quantities against \MeanScatter~dex using mean per-pixel quantities.} \rev{The fitted slope, however, differs.} \rev{With total quantities it approaches unity (1.18 against 1.39), which is what would be expected if the relation were dominated by the size of the region rather than by the coupling between emission and flux density: a larger region contains proportionally more of both.} \rev{Mean per-pixel quantities suppress this first-order dependence on area.} \rev{We therefore adopt them because they isolate the quantity of interest, not because they reduce the scatter.} 

Next, we examine the sensitivity of the calibration to the choice of disk-centre window by repeating the fit using $\pm 2$ to $\pm 5$-day intervals around central meridian passage. \rev{The results are given in the last block of Table~\ref{tab:fsys}.} \rev{The intrinsic scatter barely changes over $\pm 2$ to $\pm 4$~days and so does not discriminate between them, but the error on the two later passages grows steadily with window width, from 0.082~dex at $\pm 2$~days to 0.116~dex at $\pm 5$~days, and the slope flattens correspondingly as more limb data enters.} \rev{We adopt $\pm 3$~days: it matches the predictive performance of the narrowest window while retaining about 50\% more samples.}  We then evaluate the sensitivity of the far-side reconstruction to the intensity normalization by comparing one-sided anchoring at the ingress and egress boundaries and two-sided anchoring using the geometric mean. \rev{The absolute normalization of the inferred flux varies by \AnchorSpread\% depending on which boundary sets the scale.} \rev{This is consistent with the boundary mismatches of Table~\ref{tab:scale}: the two ends of an interval do not call for the same scale factor, so no single $S$ can satisfy both, and the absolute level of the reconstruction is correspondingly uncertain at the several tens of per cent level.} 

\begin{table}[tbp]
\centering
\tabularfont
\caption{\rev{Dependence of the calibration on the adopted uncertainty model and on the treatment of the line-of-sight projection.}}
\label{tab:fsys}
\begin{tabular}{lccc}
\topline
\textbf{Case} & \boldmath$C$ & \boldmath$D$ & \textbf{RMS (dex)}\\
\midline
\textit{WLS (adopted)} & & &\\
\quad $f_{\mathrm{sys}}=0.04$, 0.08, 0.16 & 17.389 & 1.392 & 0.083\\
\quad $\sigma_{\mathrm{pix}}=2.5$, 5, 10  & 17.389 & 1.392 & 0.083\\[3pt]
\textit{ODR} & & &\\
\quad $f_{\mathrm{sys}}=0.04$     & 17.399 & 1.816 & 0.124\\
\quad $f_{\mathrm{sys}}=0.08$     & 17.395 & 1.599 & 0.122\\
\quad $f_{\mathrm{sys}}=0.16$     & 17.394 & 1.552 & 0.113\\
\quad $\sigma_{\mathrm{pix}}=2.5$ & 17.395 & 1.620 & 0.122\\
\quad $\sigma_{\mathrm{pix}}=5$   & 17.395 & 1.599 & 0.122\\
\quad $\sigma_{\mathrm{pix}}=10$  & 17.395 & 1.614 & 0.122\\[3pt]
\textit{LOS projection} & & &\\
\quad Line-of-sight flux          & 17.319 & 1.628 & 0.083\\
\quad Deprojected flux (adopted)  & 17.389 & 1.392 & 0.083\\[3pt]
\textit{Disk-centre window} & & &\\
\quad $\pm2$~d ($n=415$)            & 17.389 & 1.324 & 0.082\\
\quad $\pm3$~d ($n=605$, adopted)   & 17.389 & 1.392 & 0.083\\
\quad $\pm4$~d ($n=765$)            & 17.393 & 1.322 & 0.091\\
\quad $\pm5$~d ($n=849$)            & 17.395 & 1.144 & 0.116\\
\hline
\end{tabular}
\tablenotes{\rev{\textbf{Note.} Under weighted least squares with flux as the dependent variable, $f_{\mathrm{sys}}$ and $\sigma_{\mathrm{pix}}$ do not enter the fit weights, so a single row is given for each; the coefficients are identical across the grid by construction.} \rev{The orthogonal-distance-regression block is the meaningful test, since there the intensity uncertainty enters the fit.} \rev{In the disk-centre window block the RMS column is the error on the two later passages, which were not used in the fit; the sample size $n$ is given because it varies with the window.}}
\end{table}
\rev{The temporal evolution is unaffected, since a change of $S$ multiplies the whole interval by a constant.} \rev{We adopt two-sided anchoring as the choice that does not privilege either boundary, and we interpret only the shape of the far-side curve.} \rev{Next, we test how far the results depend on the adopted uncertainty model.} \rev{Two tests are needed, because the answer is partly a property of the estimator.} \rev{In the weighted least-squares fit of Equation~(\ref{eq:inverse}) the weights are set by Equation~(\ref{eq:sigma_phi}) alone, so $\sigma_{\mathrm{pix}}$ and $f_{\mathrm{sys}}$ cannot affect the fitted coefficients at all; they enter only the plotted intensity error bars.} \rev{Reporting that the coefficients are unchanged under this estimator would therefore be circular.} \rev{We accordingly repeat the calibration using orthogonal distance regression, which treats the uncertainties in both variables.} \rev{For this near-side fit we vary the AIA $f_{\mathrm{sys}}$ over 0.04, 0.08 and 0.16 and $\sigma_{\mathrm{pix}}$ over 2.5, 5 and 10~DN~s$^{-1}$ per pixel; only the AIA terms enter.} \rev{Separately, we vary the EUVI systematic term over 0.06, 0.12 and 0.24 when propagating the uncertainty on the far-side.} \rev{Here the intensity uncertainty does enter the fit, and the resulting spread in the coefficients and in the test-set metrics is a meaningful measure of how far the results depend on these choices.} \rev{Table~\ref{tab:fsys} lists both sets of results.} \rev{The errors-in-variables slope is steeper than the weighted least-squares slope, in the direction expected when uncertainty in the independent variable is accounted for.} \rev{Varying the AIA $f_{\mathrm{sys}}$ across this range changes the reconstructed far-side flux by \FsysEffect\%.} \rev{The EUVI term does not enter the fit nor the scaling; so varying it changes only the width of the propagated uncertainty envelope, which is \EnvLo\% narrower at $f_{\mathrm{sys}}=0.06$ and \EnvHi\% wider at $f_{\mathrm{sys}}=0.24$ than at the adopted 0.12.} \rev{This term is therefore the dominant control on how precisely the far-side flux can be quoted, even though it has no effect on its value.} \rev{We also compare the calibration obtained from the LOS flux with that obtained from the deprojected flux $|B_{\mathrm{LOS}}|/\mu$.} \rev{The deprojection raises $\bar{\Phi}$ by an amount that grows towards the edges of the disk-centre window, and changes the fitted slope accordingly; both versions are reported in Table~\ref{tab:fsys}.} \rev{We examine the calibration residuals as a function of $\mu$.} \rev{A linear fit to the residuals gives a slope of \MuResid~dex per unit $\mu$ across the range of $\mu$ retained, which is to be compared with the intrinsic scatter of the calibration, \Sint~dex.} \rev{Finally, we address the fact that the calibration samples are not independent.} \rev{An active region evolves on timescales of hours to days, so consecutive frames separated by tens of minutes are strongly correlated, and the formal parameter uncertainties of Table~\ref{tab:coeffs}, which assume independent samples, are correspondingly optimistic.} \rev{We therefore repeat the fit on $10^3$ block-bootstrap resamples, drawing contiguous blocks of one day so that the within-block correlation is preserved.} \rev{This gives $\sigma_C = \NblockC$ and $\sigma_D = \NblockD$, larger than the formal values by factors of approximately 4.2 and 3.2 for $C$ and $D$ respectively.} \rev{Taking the slope ratio, this implies an effective independent sample size of roughly \Neff\ rather than the \Ncal\ individual samples.} \rev{The slope is therefore $D = \Dfit \pm \NblockD$, and it is this uncertainty that should be carried when the relation is applied.} \rev{The quoted test-set metrics in Section~5.1 are per-sample statistics and are unaffected by this, but the coefficient uncertainties should be read as the bootstrap values.} 

\rev{Together, these tests show that the principal conclusions are robust to the choices examined, although the fitted slope depends appreciably on whether the uncertainty in intensity is treated explicitly: orthogonal distance regression returns slopes between 1.55 and 1.82 against 1.39 for weighted least squares.} \rev{We retain the weighted least-squares relation as primary because the purpose of the calibration is to predict flux from a measured intensity, and because it gives the lower error on the independent later passages (0.083~dex against 0.11--0.12~dex).}

\begin{figure*}[tp]
    \centering
    \includegraphics[width=0.97\textwidth]{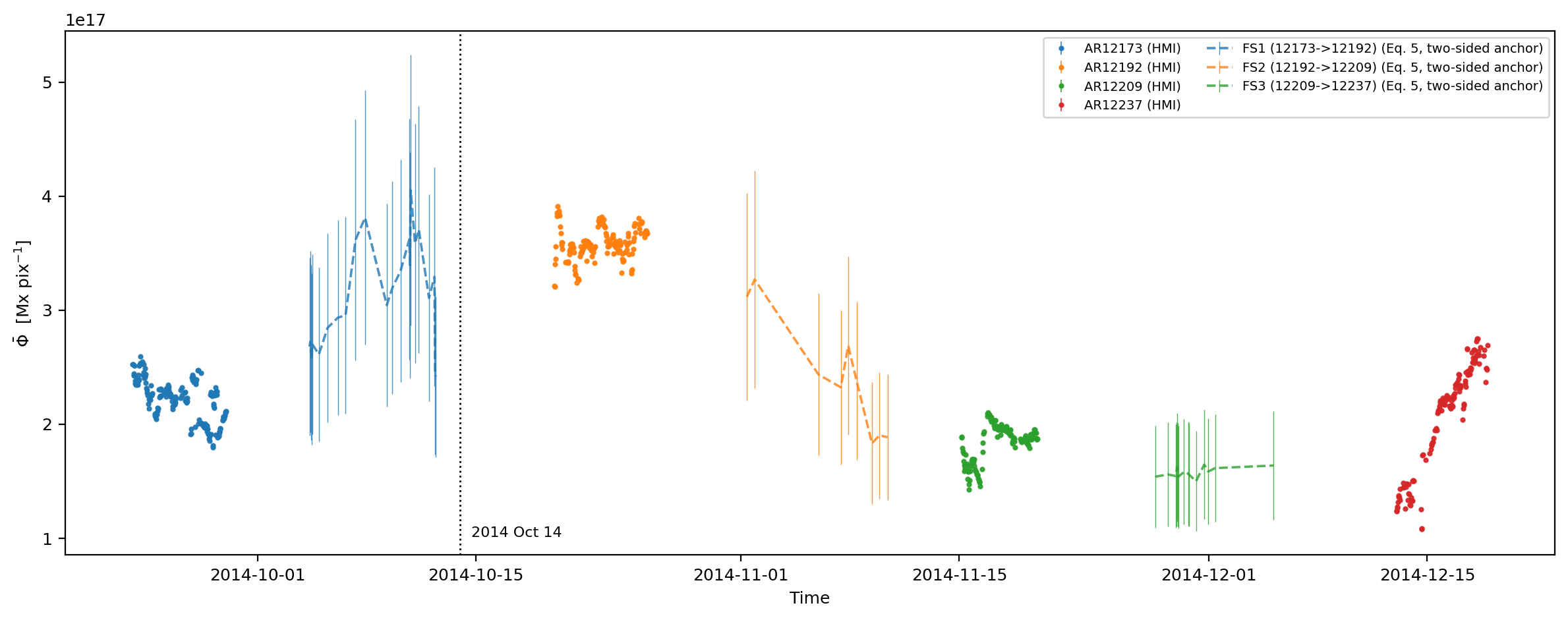}
    \caption{\rev{Stitched near-side and far-side unsigned-flux evolution of the long-lived active region across multiple rotations.} \rev{Near-side measurements are obtained from SDO/AIA and SDO/HMI, while far-side estimates come from STEREO-A/EUVI 304~\AA\ observations, using Equation~(\ref{eq:inverse}) with the EUVI intensities scaled to AIA-equivalent boundary intensities inferred from the adjacent near-side flux measurements.} The horizontal axis denotes time. The annotation marks the date of the 2014 October 14 eruptive event. \rev{The stitched curve is intended to highlight sustained multi-day evolution rather than short-timescale fluctuations within the uncertainty envelope.}}
    \label{fig:final}
\end{figure*}

\subsection{Stitched Unsigned Magnetic Flux}

Following the validation and sensitivity tests, the long-term unsigned-flux evolution of AR 12192 is visualized by stitching near-side measurements with far-side estimates using Equation~(\ref{eq:final}) to produce the continuous time-series plot in Figure~\ref{fig:final}. \rev{The three far-side intervals differ considerably in how much they constrain, and we treat them separately.} \rev{The criterion we apply is the one stated in Section~5.2: only changes that exceed the propagated uncertainty envelope are regarded as significant.}

\rev{FS2 meets that test.} \rev{The reconstructed flux declines overall from about $3.2\times10^{17}$ to $1.9\times10^{17}$~Mx~pix$^{-1}$ over roughly ten days, sampled by 9 usable EUVI measurements, a decline substantially larger than the envelope over the same interval.} \rev{This is consistent with progressive dispersal, cancellation and cumulative restructuring of the active-region field following its major eruptive activity, and it is the clearest demonstration in this work that the proxy tracks sustained far-side evolution.}

\rev{FS1 does not.} \rev{The reconstruction rises to a maximum around 10--11 October and then falls back before the 2014 October 14 eruptive event \citep{West2015,Witasse}, but the whole excursion lies within the uncertainty envelope, which spans roughly a factor of three over that interval.} \rev{We therefore make no claim of a pre-eruptive build-up from these data.} \rev{The interval contains only 23 usable EUVI samples, and the envelope is dominated by the intrinsic scatter of the calibration rather than by measurement noise.}

\rev{FS3 is flat to within its uncertainties and, with 15 usable samples, constrains the flux evolution only weakly.} \rev{The boundary offsets of Table~\ref{tab:scale} are visible directly in Figure~\ref{fig:final} as steps between the near-side measurements and the adjacent reconstructed segments.} \rev{They are a consequence of adopting one scale factor per interval, as set out in Section~4.4, and are not evidence of real flux change across the limb.} \rev{It follows that it is the shape of each far-side segment, rather than its absolute level, that carries information about the region's evolution.}

\section{Conclusions}

\rev{We have presented a method, with stated uncertainties, for deriving the far-side unsigned magnetic flux history of an active region using EUV 304~\AA\ emission as a proxy.} Through the combination of near-side calibration, spatially consistent tracking, and cross-instrument intensity anchoring, the analysis shows that EUV observations can provide a quantitative estimate of magnetic evolution even when direct magnetograms are unavailable. \rev{Testing the relation on the two later near-side passages, which were not used in fitting it, supports this: the relation retains useful accuracy beyond the data it was fitted to.} \rev{A single relation is used for both the far-side reconstruction and the boundary anchoring, and the magnetogram flux is deprojected to the radial direction before calibration.} While the method is demonstrated here for a single long-lived active region, most components of the procedure are sufficiently general to be applied more widely, including the image registration, segmentation, region selection, and fixed-area reprojection steps. \rev{Practical choices such as the segmentation multiplier $k$, the minimum-area threshold, and the disk-centre window are reasonable starting points, though they may need adjusting for other datasets.} The coefficients of Equation~(\ref{eq:inverse}), by contrast, should be recalibrated for a representative sample if the method is to be extended to other regions or different phases of the solar cycle.

At the same time, the scope of the present result should be stated clearly. The present analysis recovers unsigned flux and does not directly constrain free magnetic energy, helicity, or overlying magnetic structure. Thus, while an increase in the recovered flux may be compatible with enhanced eruptive potential, attributing any specific trend to a particular far-side event requires support from independent observations, helioseismic measurements, and/or morphology-based diagnostics. \rev{Since Equation~(\ref{eq:inverse}) carries over to near-side passages not used in the fit with an RMS scatter of \HoRMS~dex and a median fractional error of \HoPct\%, the \emph{relative} far-side variations are constrained at a comparable level.} \rev{The absolute level is not: as Table~\ref{tab:scale} shows, the two boundaries of a far-side interval imply scale factors that differ by up to 43\%, so the normalization of each far-side segment carries an uncertainty of several tens of per cent.} In this sense, the method is best viewed as a tool for recovering sustained, large-scale far-side unsigned-flux evolution rather than short-timescale fluctuations within the uncertainty envelope. Accordingly, only changes that exceed this uncertainty envelope, particularly sustained multi-day increases or decreases, should be regarded as physically significant. \rev{Applying that criterion to the present data, the sustained decline recovered during FS2 qualifies; the excursions recovered during FS1 and FS3 do not.} A particularly important next step is to apply this framework to a broader sample of long-lived active regions and use the resulting far-side unsigned-flux histories for science interpretation of pre-eruptive evolution and for data-driven modelling of historical and archival EUV observations.

\section{Acknowledgements}

The authors thank the Space Astronomy Group at the U. R. Rao Satellite Centre, Indian Space Research Organisation, Bengaluru, for support and facilities. SDO is a mission of NASA's Living With a Star programme. STEREO is a mission in NASA's Solar Terrestrial Probes programme. Data were accessed via JSOC.

\end{document}